%% file: SousaARESMOOG.tex
\documentclass[graybox]{svmult}

\usepackage{mathptmx}       
\usepackage{helvet}         
\usepackage{courier}        
\usepackage{type1cm}        
\usepackage{makeidx}         
\usepackage{graphicx}        
\usepackage{multicol}        
\usepackage[bottom]{footmisc}

\usepackage{bm}

\makeatletter
\def\verbatim{\small\@verbatim \frenchspacing\@vobeyspaces \@xverbatim}
\makeatother

\makeindex             

\begin{document}

\title*{Measuring Stellar Atmospheric Parameters with ARES+MOOG}
\author{S. G. Sousa and D. T. Andreasen}
\institute{S. G. Sousa \at Instituto de Astrof\'isica e Ci\^encias do Espa\c{c}o, Universidade do Porto, CAUP, Rua das Estrelas, 4150-762 Porto, Portugal, \email{sergio.sousa@astro.up.pt}
\and D. T. Andreasen \at Instituto de Astrof\'isica e Ci\^encias do Espa\c{c}o, Universidade do Porto, CAUP, Rua das Estrelas, 4150-762 Porto, Portugal, \email{sergio.sousa@astro.up.pt}}
%
%
\maketitle

\abstract*{The technical aspects in the use of an Equivalent 
Width (EW) method are described for the derivation of spectroscopic stellar parameters 
with ARES$+$MOOG. While the science description behind this method can 
be found in many references, here the goal is to provide a user manual 
approach for the codes and scripts presented in for the tutorial. All the 
required data is available online\footnote{https://github.com/sousasag/school\_codes}.
}

\abstract{The technical aspects in the use of an Equivalent 
Width (EW) method are described for the derivation of spectroscopic stellar parameters 
with ARES$+$MOOG. While the science description behind this method can 
be found in many references, here the goal is to provide a user manual 
approach for the codes and scripts presented in for the tutorial. All the 
required data is available online\footnote{https://github.com/sousasag/school\_codes}.
}

\section{Introduction}
\label{sec:1}

Several methods are used for the derivation of stellar spectroscopic 
parameters. These can be divided into two main groups. One based on 
spectral synthesis for which synthetic spectra are created and compared 
with the observed spectrum to find the best fit. The second group adopts 
a line by line analysis strategy, measuring the strength of observed spectral 
lines and then estimating abundances and applying Physics criteria such 
as the ionization and excitation balance to find the best 
spectroscopic stellar parameters for the observations. The description 
of the physics behind these methods can be found in many books\cite{gray-book}.

A brief work-flow for an EW method can be described as follows: For a high 
quality observed stellar spectrum, starting by measuring the strength of 
several spectral lines (e.g. EWs with ARES). These measurements are then converted 
into individual line abundances which are computed (e.g. with MOOG) using 
stellar atmospheric models, normally assuming Local Thermodynamical 
Equilibrium (LTE) approximation. The model parameters are then adjusted until
the individual line abundances show evidence of the excitation and ionization 
balance. This step can be automatized if a proper minimization method is adopted.

The rest of the document will focus on the technical aspects of the 
codes provided for this tutorial and how a student can use them to actually 
derive spectroscopic stellar parameters. A more specific description of the 
ARES$+$MOOG method can be found in \cite{sousa-2014}. The next sections will
be based on the codes and test data available online.

\section{Requirements}
\label{sec:2}

This tutorial was implemented to be executed on a Linux machine. Note that even 
without an access to a Linux machine, today it is very easy to emulate one by 
using virtual machines (e.g. VirtualBox). All the instructions provided both 
in this document and in the repository are compatible for Debian/Ubuntu systems, 
but can be easily adapted for other Linux flavors.

\subsection{Main codes for the tutorial}

\paragraph{\textbf{ARES}}

ARES is a C code that allows a fast and automatic measurement of EWs of spectral 
absorption lines. The ARES code is a submodule in the \textit{school\_codes} 
repository and therefore easily updated with the root ARES repository\footnote{https://github.com/sousasag/ARES} 
which contains more information. For a more detailed description of 
the code please see the ARES papers \cite{sousa-2007, sousa-2015}. 
Note that the latest version of ARES is already able to deal with in-situ radial 
velocity correction and automatic parameterization for the continuum level which 
makes the spectral analysis easier and consistent. In order to 
compile ARES it is required some external libraries (e.g. CFITSIO, GSL, etc.) which 
can be easily installed in Linux machines. The compilation of the code is then handled 
by the Makefile provided in the repository.

\paragraph{\textbf{MOOG}}

MOOG is a code that performs a variety of LTE spectral analysis. The original 
code can also be found in its homepage\footnote{http://www.as.utexas.edu/$\sim$chris/moog.html}. 
The code available in the \textit{school\_codes} repository is an adapted 
version of MOOG2014 modified to neglect its (non-free) plot library dependency. 
In this tutorial it is only used the silent version of the code to make 
abundance computations for our EW method. The compilation of the code is also 
handled by the Makefile.

\subsection{Spectral test data}

\paragraph{\textbf{Observed Spectrum}}

The spectra for the analysis should be of good quality, in terms of resolution 
(R $>$ 30000) and signal-to-noise ratio (S/N $>$ 100). Three HARPS 
spectra are provided for the tutorial. These spectra have R $\sim$ 110000, and S/N $>$ 300. 
These spectra is provided in the folder \textit{spectra} of the repository. 
These are standard FITS 1D spectra. In order to be compatible with ARES they 
need to have the standard keywords CDELTA1 and CRVAL1 and the units of the 
wavelengths should be Angstrom. Alternatively ARES can also read ASCII spectra 
where the format must be two columns, the first with the wavelengths and the 
second with the flux. In this tutorial only the standard 1D fits 
files are used.

\paragraph{\textbf{Linelist}}

The strength of specific absorption spectral lines should be measured in this 
analysis. As in most EW methods, iron lines are mostly used because they strongly 
populate the solar-type spectrum and because iron can be used as a proxy for 
the stellar metallicity. The linelist provided online (\textit{iron lines\_parameters.dat}) 
compiles 263 Fe I and 36 Fe II lines. These lines were carefully selected 
\cite{sousa-2008} and its atomic data were revised using a solar spectrum as 
a reference making it possible to use of a differential analysis \cite{sousa-2014}. 
This linelist is used both for ARES and MOOG. For ARES the important information 
is only the wavelength of the lines. The format in this case is to keep the first 
column as wavelength in Angstrom. For the individual abundance computation MOOG 
needs the atomic data, the Excitation Potential (E.P.), and oscillator strengths 
(log gf), the 2nd and 3rd columns respectively. The format of this file should 
be kept fixed so the scripts work without a problem.

\subsection{Script Codes}

\paragraph{\textbf{Input linelist for MOOG}}

The EW measurements and the atomic data for the linelist require a special format 
to be used by MOOG. To facilitate this task it is provided a python script 
(\textit{make\_moog\_lines.py}) that reads the atomic data from the linelist and 
the output file from ARES to compile the required formated input file for MOOG.

\paragraph{\textbf{Creating a stellar atmospheric model}}

For the computation of the individual line abundance it is required the use an 
atmospheric stellar model. MARCS models\footnote{marcs.astro.uu.se} are used in this tutorial.
For this task it is provided a script that interpolates a 
grid of MARCS models for specific stellar parameters. The folder \textit{interpol\_models\_marcs} 
in the repository includes a script for this task. In order to make it work, it is 
first required to follow the instructions to download and extract the grid of MARCS 
models. The use of the script to get a specific atmospheric stellar model is straightforward 
and it will be described later in this document.

\paragraph{\textbf{Excitation and Ionization Correlation}}

To understand how the individual line abundances depend on the stellar atmospheric 
parameters and how to find the best stellar parameters, it is provided a plotting script 
that allows the visualization of the correlations/indicators needed to constrain the parameters. 
These include the abundance vs. E.P., abundance vs. Reduced EW., the information 
on $|<FeI> - <FeII>|$ and the average [Fe/H] compared with the model [M/H]. For more details
about these indicators see section \ref{sec:5} and \cite{sousa-2014}. The python 
script (\textit{running\_dir/read\_moog\_plot.py}) 
allows to plot these correlations as well as display the information of these 
indicators as it is presented later in Fig. \ref{fig:1} and \ref{fig:2}.

\section{Step by Step Tutorial}
\label{sec:3}

\begin{figure}[t]
\centering
\includegraphics[scale=.5]{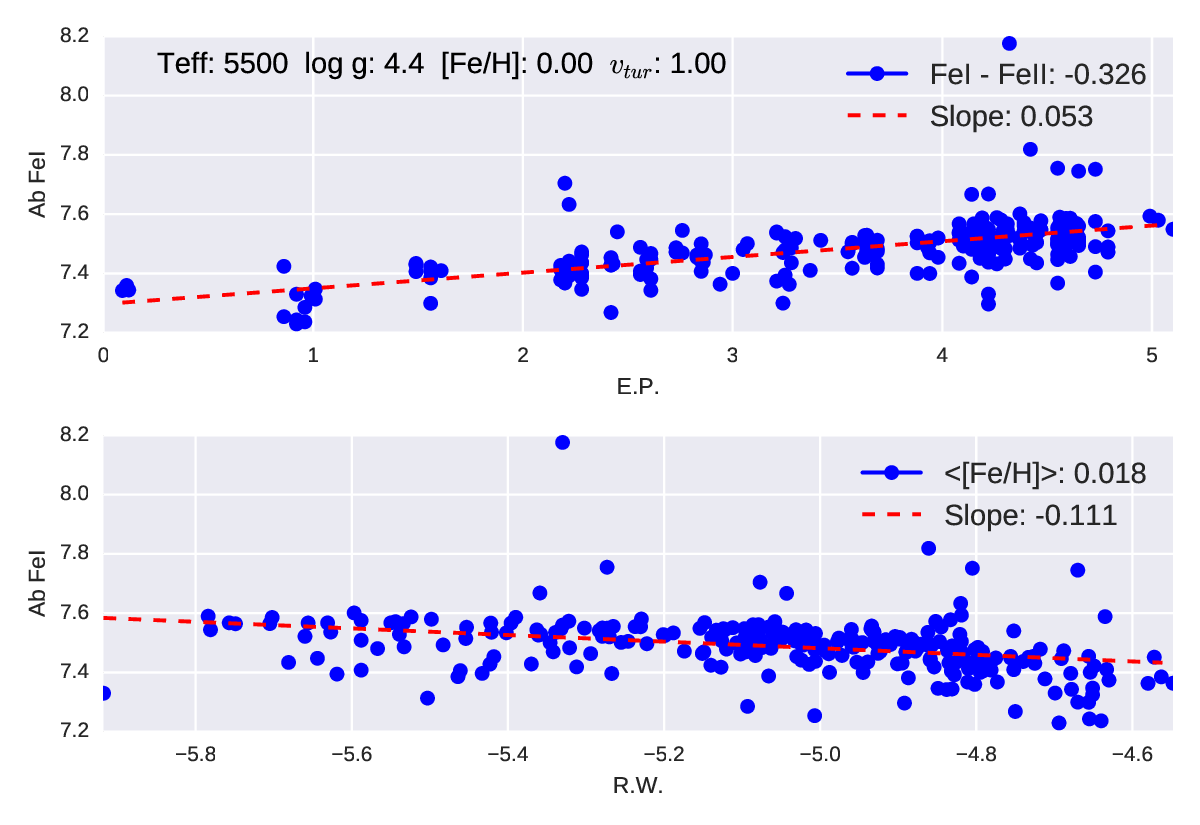}
\caption{The result of the first iteration of the \textit{TestA.fits} spectrum.}
\label{fig:1}       
\end{figure}

The criteria to derive good spectroscopic parameters with this method rely 
on the model that enforces the excitation and ionization balance 
for the line abundances. Therefore the final step of this iterative method is 
when it is found the same abundance for all the lines which will translate in zero 
slopes for abundance vs. E.P., abundance vs. Reduced EW. and $<FeI> = <FeII>$. 
How to reach this? The steps to derive stellar parameters, where 
the spectrum \textit{TestA.fits} is used for this example, are described as follow:

\begin{enumerate}

 \item {\textbf{Location}: To start the process lets first define 
 a reference folder. Let the folder \textit{running\_dir} be our reference in order 
 to run the codes and scripts in this tutorial. To confirm this, when displaying 
 the spectra available in the repository the result should be:
 \begin{verbatim}
 running_dir$ ls ../spectra/
 TestA.fits  TestB.fits  TestC.fits
 \end{verbatim}
 }
 
 \item {\textbf{Measuring EWs}: In this folder it is already present the required ARES 
 input file (\textit{mine.opt}) containing the recommended parameters for 
 high quality spectra. Details for these parameters may be found in \cite{sousa-2007,sousa-2014}. 
 Running ARES will display plenty of information in the terminal. In the end the 
 output file (\textit{TestA.ares}) will be created as well as a log file (\textit{logARES.txt}) 
 containing all the relevant information. To run ARES for the \textit{TestA.fits} 
 spectrum\footnote{Note that the \textit{mine.opt} file should be adapted for the other test spectrum.}:
 \begin{verbatim}
 running_dir$ ../ARES/./ARES
 \end{verbatim}
 }
 
 \item{\textbf{MOOG input linelist}: To create the line list with the EWs 
 in the correct format it should be provided the ares output file 
 (\textit{TestA.ares}) as well as the linelist with the reference atomic 
 data (\textit{ironlines\_parameters.dat}). This can be done with:
 \begin{verbatim}
 running_dir$ .././make_moog_lines.py TestA.ares ../ironlines_p
 arameters.dat 
 Saved in: lines.TestA.ares
 \end{verbatim}
 }
 
 \item{\textbf{Loop start}: The beginning of the iteration loop is here where 
 it is created a specific MARCS model. Starting the loop with parameters that 
 represent an average dwarf solar type star (Teff: 5500 K, logg: 4.40 dex, 
 $[M/H]$ = 0 dex, $v_{tur} = 1.0$ km/s), to generate this model (creating 
 the file \textit{out\_marcs.atm}):
 \begin{verbatim}
 running_dir$ ../interpol_models_marcs/./make_model_marcs.bash 
 5500 4.4 0.0 1.0
 \end{verbatim}
 }
 
 \item{\textbf{MOOG}: The next iterative step is to compute the individual 
 line abundances assuming the stellar atmospheric model. For this step MOOG 
 will use an already defined input file (\textit{batch.par}) and will then 
 create an output file with the computed line abundances (\textit{output.moog}). 
 To run MOOG in its silent mode:
 \begin{verbatim}
 running_dir$ ../MOOG2014/./MOOGSILENT
 \end{verbatim}
 }
 
 \item{\textbf{Control check}: This is the crucial step where it is validated 
 the stellar parameters for the model. Again, good parameters are found when 
 the same abundance is derived for all the lines. The slopes should therefore 
 be negligible (all indicators having values $<$ 0.005 is a good criteria). If 
 significant  correlations are found in these indicators, these can be used 
 to adjust the parameters back in step 4. To check the status of these indicators 
 the plotting python script can be used:
 \begin{verbatim}
 running_dir$ python read_moog_plot.py output.moog 
 Model Parameters: Teff logg vtur [M/H]
                   5500 4.4 1.00 0.00
 -----------------------------
 |   Slope  E.P. :0.053
 |   Slope  R.W. :-0.111
 |  Fe I  - Fe II:-0.326
 |      [Fe/H]   :0.018
 | [FE/H] - [M/H]:0.018

\end{verbatim} 
 
 Figure \ref{fig:1} shows the plots that appear in the last step. The 
 indicators values for this first iteration are far from zero. Therefore 
 it is require to go back to step 4 and try different parameters. The strategy 
 to minimize this problem and change the parameters in the correct direction is 
 discussed in the next section.
 }
 
\end{enumerate}

\section{The strategy to find the best parameters}
\label{sec:4}

In this section it is briefly described the technical strategy to find the best 
model. A complete description of this strategy is found in \cite{sousa-2014}. 
From figure \ref{fig:1} the four indicators values provide the necessary 
hints to reach a better model for the next iteration.

\begin{itemize}
 \item {\textbf{Slope E.P.}: This indicator is the one that strongly depends on 
 the temperature of the model. Negative values of this slope means that the correct 
 temperature should be lower. To fix positive values the temperature should 
 increase in the next iteration. This indicator controls the excitation balance. 
 Given the very high number of iron lines, the temperature is one of the best 
 constrained parameters with this method.
 }
 
 \item {\textbf{Slope R.W.}: This indicator is connected with the microturbulance 
 parameter which basically controls the abundance determination for the stronger 
 lines where saturation becomes specially significant in the wings of the absorption 
 lines\cite{gray-book}. For negative slopes the microturbulance should be reduced, 
 while positive values it means the microturbulance is overestimated.
 }
 
 \item {$\bm{<FeI> - <FeII>}$: Since FeII lines are more sensitive to surface gravity 
 than FeI lines\cite{gray-book}, this indicator can be used to control the log g 
 of the  model. For negative values then log g should decrease for the next iteration, 
 while for positive values log g should be higher.
 }
 
 \item {$\bm{[FE/H] - [M/H]}$: More than an indicator, this is actually a logical 
 physical constraint for the model. Meaning that the input metallicity should output 
 a compatible global iron abundance. Here it is assumed that the iron abundance 
 is a proxy for the stellar metallicity. Therefore if the iron abundance 
 is larger than the model metallicity (positive indicator) then it
 should be increased for the next iteration. If the indicator 
 is negative this means that the metallicity of the model is overestimated.
 }

 \end{itemize}
 
\begin{figure}[t]
\centering
\includegraphics[scale=.5]{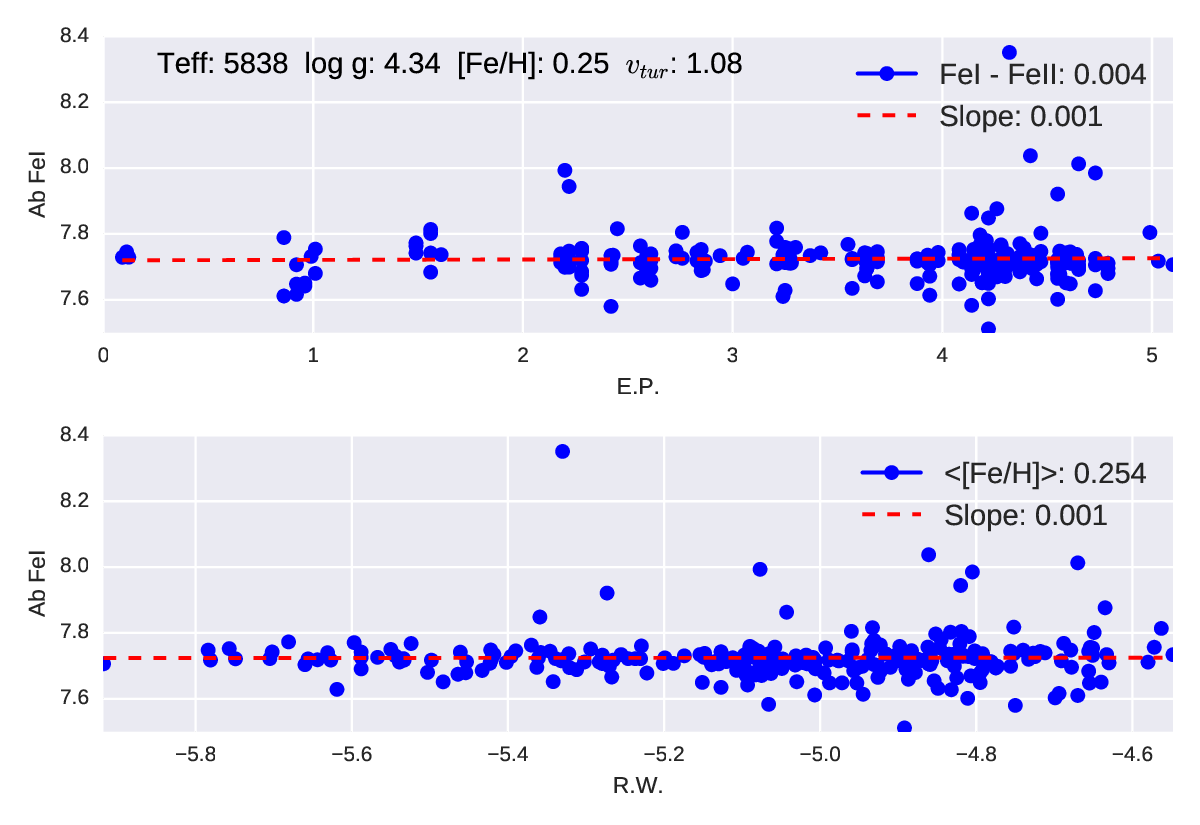}
\caption{A good model dericed for the \textit{TestA.fits} spectrum.}
\label{fig:2}       
\end{figure} 
 
These indicators can be used to find a better model for the spectrum \textit{TestA.fits}. 
Figure \ref{fig:2} shows one good 
result for a model with Teff: 5838 K, logg: 4.34 dex, $[M/H]$ = 0.25, $v_{tur} = 1.08$ 
km/s. With this model all the indicators are close to zero (all $<$ 0.005). These results 
can still improve, given the clear presence of outliers in the figure. In this case 
most of these outliers are overestimating the iron abundance (compared with the average). The 
cause for this may be related with less reliable, and overestimated, EW measurements for 
those specific spectral lines and therefore removing these outliers will improve the result. 

For this tutorial, it is provided in table \ref{tab:1} the identification of the stars for 
each test spectrum provided. The table also summarizes the stellar parameters which were 
derived by using the same method with remotion of outliers. For the spectrum \textit{TestA.fits}, 
which correspond to HD128620 (also known as $\alpha$ Cen A) the results are quite close to 
the ones presented in Fig. \ref{fig:2}. Note that temperature is within 10 K, and the metallicity 
should be slightly lower (by only 0.02 dex) as expected given that the outliers overestimate 
the iron abundance. Although here it is not discussed any error analysis for this 
method, it is clear from the results that the precision of the method is quite high, just by 
considering the very low spread of the parameters derived without the lines with problematic 
EW measurements.

\begin{table}
\label{tab:1}       
\centering
\caption{Parameters derived with the ARES+MOOG method for the provided test spectra.}

\begin{tabular}{p{2cm}p{2cm}p{1.5cm}p{1.5cm}p{1.5cm}p{1.5cm}}
\hline\noalign{\smallskip}
File & Star     & Teff & log g & [Fe/H] & $v_{tur}$ \\
     &         & (K)  & (dex) & (dex)  &  (km/s)   \\
\noalign{\smallskip}\svhline\noalign{\smallskip}
TestA.fits & HD128620 & 5832 &  4.33 &  0.23  &   1.11    \\
TestB.fits & HD128621 & 5234 &  4.40 &  0.16  &   0.90    \\
TestC.fits & HD179949 & 6287 &  4.54 &  0.21  &   1.36    \\
\noalign{\smallskip}\hline\noalign{\smallskip}
\end{tabular}
\end{table}

\section{Summary}
\label{sec:5}

The technical aspects were presented for running the tutorial 
to derive spectroscopic stellar parameters with the ARES+MOOG EW method. 
The strategy for finding a good model based on the indicators provided for 
each iteration is briefly described. This document is a complement 
for the material provided online. For a more complete understanding of 
the ARES+MOOG method it is strongly advised the reading of previous 
works\cite{sousa-2007, sousa-2015, sousa-2014}.

\begin{acknowledgement}
"S.G.S and D.T.A. acknowledges the support by Fundação para a Ciência e 
Tecnologia (FCT) through national funds and a research grant (project 
ref. UID/FIS/04434/2013, PTDC/FIS-AST/7073/2014, and grant ref: 
CAUP-09/2014-BD). S.G.S. also acknowledge the support from FCT through 
Investigador FCT contract of reference IF/00028/2014 and POPH/FSE (EC) 
by FEDER funding through the program “Programa Operacional de Factores 
de Competitividade − COMPETE”.
\end{acknowledgement}

\input{referenc}

\end{document}

%% file: referenc.tex
%
%
%